\documentclass[12pt,preprint,graphics]{aastex}
\def\etal{{\it et al. }}
\def\asec{$^{\prime\prime }$}
\def\amin{$^\prime $}
\def\deg{\hbox{$^\circ$}}

\shortauthors{Wilking et al.}
\shorttitle{Brown Dwarf Candidates}

\begin{document}

\title{Low Mass Stars and Substellar Objects in the NGC 1333 Molecular Cloud}

\author{Bruce A. Wilking \altaffilmark{1}}
\affil{Department of Physics and Astronomy, University of Missouri-St. Louis\\
8001 Natural Bridge Road, St.  Louis, MO 63121\\ bwilking@umsl.edu}

\author{Michael R. Meyer} 
\affil{Steward Observatory, The University of Arizona, Tucson, AZ 85721 \\
mmeyer@gould.as.arizona.edu}

\author{Thomas P. Greene \altaffilmark{1}}
\affil{NASA/Ames Research Center, M.S. 245-6 \\
Moffett Field, CA  94035-1000 \\
tgreene@mail.arc.nasa.gov}

\author{Ayman Mikhail \altaffilmark{1}}
\affil{Department of Physics and Astronomy, University of Missouri-St. Louis\\
8001 Natural Bridge Road, St.  Louis, MO 63121}

\and

\author{Glenn Carlson}
\affil{Department of Physics and Astronomy, University of Missouri-St. Louis\\
8001 Natural Bridge Road, St.  Louis, MO 63121 \\
gcarlson@mail.win.org}

\altaffiltext{1}{Visiting Astronomer at the Infrared Telescope Facility,
which is operated by the University of Hawaii under contract from the
National Aeronautics and Space Administration.}

\pagebreak
\begin{abstract}
\setcounter{page}{2}
We present the results of near-infrared imaging and low-resolution 
near-infrared spectroscopy of low mass objects in the NGC 1333 molecular cloud.
A JHK survey of an 11.4\amin x 11.7\amin\ area of the northern cluster was conducted
to a sensitivity of K$\le$16 mag.  Using near-infrared magnitudes and colors from this
and previously published surveys, twenty-five 
brown dwarf candidates were selected toward the high extinction cloud core.
Spectra in the K band were obtained and comparisons of the depths of water vapor
absorption bands in our candidate objects with a grid of dwarf,
subgiant, and giant standards were made to derive spectral types. 
These data were then used to derive effective
temperatures and stellar luminosities which, when combined with
theoretical tracks and isochrones for pre-main sequence objects,
resulted in estimates for their masses and ages.  The models suggest a
median age for the sample of $<$1 Myr with substellar masses for at least 
9 of the candidates including the x-ray flare source ASR~24.  Surface gravities
have been estimated for the brown dwarf candidates and, for a given spectral type,
found to resemble more
closely dwarfs than giants.
Using the near-infrared imaging data and age estimates from the spectroscopic sample, 
an extinction-limited sample in the northern cluster was defined.  Consistent with 
recent studies of other young clusters, this sample exhibits an accretion
disk frequency of 0.75$\pm$0.20 and a mass spectrum slope across the hydrogen-burning limit
of $\alpha \le$1.6 where dN/dM $\propto$ M$^{-\alpha}$.   

\end{abstract}

\section{Introduction}

The discovery and characterization of brown dwarfs in young clusters can help answer
fundamental questions about the star formation process.  
Do brown dwarfs dominate in number or mass over low mass stars in young clusters?
Do the mass functions of young clusters exhibit a characteristic mass and do they
resemble the field star Initial Mass Function (IMF)?
Are there 
differences between the infrared or x-ray properties of young low mass stars and brown dwarfs that
would suggest separate formation mechanisms? 
For example, it has been suggested
that some brown dwarfs are ejected from young, multiple star systems thus limiting
their accretion phase (Reipurth \& Clarke 2001).
Studies of the mass functions of young clusters rely critically on the substellar population
to define the peaks of the mass functions and possible low-mass cut-offs (e.g., 
Hillenbrand \& Carpenter 2000; Luhman \etal 2000; B\'ejar \etal 2001; Najita \etal 2000;
Muench \etal 2002).
Such studies will not only help establish whether
there are significant variations in the mass functions between young clusters but if
these depend on the physical conditions of the cluster gas (Briceno \etal 2002).

Young brown dwarfs are readily observed in near-infrared surveys of dark clouds
since their luminosity is derived from gravitational contraction as well as a brief
period of deuterium burning 
(e.g., Burrows \etal 1997).
For example, models predict that in the nearest star-forming regions an object with a
mass of 0.04 M$_{\odot}$ will have a K magnitude of 12.4 at an age
of 1 Myr (D'Antona \& Mazzitelli 1998).  Recent infrared surveys of young clusters such as
the Orion Nebula Cluster (ONC) and IC 348 suggest sizable populations of brown dwarfs
(e.g. Hillenbrand \& Carpenter 2000; Luhman \etal 2000; Muench \etal 2002;
Muench \etal 2003).  However due to an ambiguity in the derived age and mass, mass estimates 
from infrared photometric data alone are reliable only in a statistical sense.
Determination of the
effective temperature through spectroscopy removes this ambiguity and leads 
to more accurate determinations of the visual extinction and stellar luminosity,
and with the help of models, the  
age and mass.  Hence, potential substellar objects can be identified and studied 
on an individual basis.
Infrared spectroscopy is ideal for confirming the nature of young brown dwarf candidates
because of the obscuration by dust inherent in star-forming regions and the presence of
broad water vapor absorption bands in cool photospheres that can be used to derive
effective temperatures using low resolution (R$\sim$300) spectra (e.g., Wilking, Greene,
\& Meyer 1999, hereafter WGM).

The NGC 1333 reflection nebula and its associated dark cloud L1450 are
part of a chain of molecular clouds in the Perseus region (e.g., Sargent
1979; Loren 1976).  Analysis of Hipparcos data suggests a distance of 300
pc (de Zeeuw \etal 1999; Belikov \etal 2002).  The observations of
emission-line stars and Herbig-Haro objects first established NGC 1333 as
an active region of star formation (Herbig \& Rao 1972; Herbig 1974).
Surveys of the cloud at near-infrared, far-infrared, and submillimeter
wavelengths have revealed that it is forming primarily low mass stars with the 
most massive young stars, SVS~3 and BD+30\deg\ 549, exhibiting spectral types 
in the range of B6--B9
(e.g., Strom,
Vrba, \& Strom 1976; Harvey, Wilking, \& Joy 1984;
Aspin \etal 1994; Lada, Alves, \& Lada 1996; Jennings \etal 1987;
Sandell \& Knee 2001, Aspin 2003).  Most of the young stellar objects (YSOs) in the NGC 1333
cloud are concentrated in the highest column density gas and are arranged into a
northern and southern cluster, each with about 70 members.

We utilized deep (K$\le$16) near-infrared surveys of the NGC 1333 cloud core,
including our new survey of the northern cluster,
to select 25 brown dwarf candidates among the low mass population. 
Infrared spectra of the candidates were obtained, in addition to a collection 
of dwarf, subgiant, and giant
standards, and a water
vapor index was developed to yield 
estimates for their effective temperatures assuming dwarf to subgiant surface gravities.
Using theoretical tracks and isochrones, we used our derived effective temperatures
and stellar luminosities to identify brown dwarfs and to investigate their collective properties
such as mass, age, infrared excess, x-ray emission, and surface gravity.
Finally, we use our survey of the northern cluster and the median age of the
brown dwarf candidates to assemble an extinction-limited sample of YSOs in the NGC 1333 core.
We use this sample to explore the stellar density, disk frequency, and the broad features
of the mass spectrum for comparsion with other young clusters and the field star
IMF.

\section{Observations and Reduction}
To assist in the identification of brown dwarf candidates, we first obtained
a new, deep near-infrared survey of the northern cluster.  These data were
supplemented with published infrared surveys of the southern cluster
to construct a sample of candidates for near-infrared
spectroscopy.  

\subsection{A New Infrared Survey of the Northern Cluster}

Near-infrared images of the NGC~1333 northern cluster were obtained using
the 1.6 meter 
telescope at Mt. Bigelow (MBO) near Tucson, AZ on the nights of
1999 October 21-23.
The University of Arizona's 256x256 HgCdTe Infrared Imager
was used at wavelengths of J (1.25 $\mu$m), 
H (1.6 $\mu$m),
and K (2.2 $\mu$m) 
with a pixel scale of 0.92\asec\ pixel$^{-1}$ (see Rieke \etal 1993 for a description of
the instrument).

The northern cluster was mapped using a 5x5 raster pattern with half-frame
overlap.  The total area covered was 11.4\amin x 11.7 \amin.
The southwest corner of the mosaic is located at RA(J2000)=
3$^h$ 28$^m$ 40.8$^s$, DEC(J2000)=+31\deg\ 18\amin\ 19\asec\ and the
northeast corner at  RA(J2000) = 3$^h$ 29$^m$ 41.4$^s$, DEC(J2000)=
+31\deg\ 29\amin\ 56\asec.  The boundaries of the survey relative to the
high column density gas in the core are shown in Fig. 1.
Integration times per frame were 30 sec, 20 sec, and 10 sec at J, H, and
K, respectively.
Individual frames were sky-subtracted using 
a median-combined image from a subset of frames within the 5x5 raster
that was free of bright stars.
The same median-combined image, with dark current subtracted,
was used to flat-field frames.
Mosaics were constructed for each 5x5 raster using the routine SQMOS 
within the Infrared Reduction and Analysis Facility (IRAF)  
\footnote{IRAF is distributed by the National Optical Astronomy Observatories,
which are operated by the Association of Universities for Research
in Astronomy, Inc., under cooperative agreement with the National
Science Foundation.}.  
We repeated the 5x5 raster 10 times at J and K, and 8 times at H.  The mosaics
at a given wavelength were combined to yield a total integration time of 
20, 10.7, and 6.7 minutes at J, H, and K, respectively, in the central 
regions of the field.
 
Automated source extraction was performed on the final K mosaic
using DAOFIND with a threshold of 5 times the rms noise.
A total of 213 unsaturated sources were extracted at K.
The pixel positions of these
sources were used to center apertures for photometry of individual sources
in all three bands resulting in a total of 211 sources measured at H
and 187 at J.
Aperture photometry was performed using APPHOT in IRAF using an aperture
diameter of 3.7\asec.  This aperture was twice the average FWHM
of the stellar images and provided the optimum signal-to-noise (Howell 1989). 
Observations of faint UKIRT standards were used to determine aperture corrections of 
0.28, 0.29, and 0.22 mag at J, H, and K, respectively.

The JHK data are presented in Table 1 in the CIT photometric system.  
To place our photometry (labeled MBO) in the CIT photometric system, we first
compared our magnitudes for 54 sources with K$\lesssim$13.5 mag with those 
determined from the SQIID survey of Lada, Alves, \& Lada (1996, hereafter LAL).
To place our H and K data in the same photometric system as LAL, only a 
zero point correction determined from this comparison was needed.
The conversion of our J magnitudes used the following
correction:
\begin{equation}
(J-K)_{SQIID} = 1.03(J-K)_{MBO} - 0.14.
\end{equation}
To convert our photometry
into the CIT photometric system, which is very close to the SQIID system, 
we used the prescription of Kenyon
\etal (1998).  Only the statistical errors in the photometry
are quoted in Table 1.  However these errors can be exceeded by
uncorrected variations in the responsivity
across the array due to vignetting
that introduce an uncertainty
of $\pm$0.05 mag.  Errors as large as
$\pm$0.10 mag can occur in the outer regions of the mosaic due to the lower signal-to-noise.
A histogram of the number of sources vs. apparent magnitude was constructed for each wavelength
and the turnovers in the histograms were used to estimate
the completeness limits of our survey at J, H, and K to be
17.5, 16.5, and 16 mag, respectively.
The positions
in Table 1 were computed relative to a secondary reference frame defined by
Herbig \& Jones (1983) and are accurate to $\pm$0.5\asec.

\subsection{Selection of Brown Dwarf Candidates}

Brown dwarf candidates were selected based upon their broadband
near-infrared colors.  The majority of the 25 candidates lie in a region
of the K vs. (H-K) diagram shown in Fig. 2 with M$<$0.1 M$_{\sun}$,
and K$<$14 mag.  We adopted an age of 1 Myr based on the K luminosity
function by LAL.
The isochrones in Fig. 2
were derived from the models of
D'Antona \& Mazzitelli (1994, 1997, M$>$0.2 M$_{\sun}$) and D'Antona \&
Mazzitelli (1998, M$\le$0.2 M$_{\sun}$) in the CIT photometric
system using intrinsic colors and bolometric corrections for
dwarf stars (see Appendix A in WGM99).
The data
and reddening law (Cohen \etal 1981) are also in the CIT photometric
system.  
The list of brown dwarf candidates observed spectroscopically
is given in Table 2.

\subsection{Low Resolution Spectroscopy}

Infrared spectroscopic observations were obtained using the 3-m
NASA Infrared Telescope Facility at Mauna Kea, Hawaii in 2000
November 10-13.  Spectra in the K band were obtained for
25 brown dwarf candidates, eight M giants, and eight M subgiants using the
256 x 256 InSb facility infrared camera (NSFCAM) with the HKL
grism and a 0.3\asec\ pixel$^{-1}$ scale.
The 0.6\asec\ slit allowed us to cover the 2.0 $\mu$m--2.5 $\mu$m band
with a resolution of R = $\lambda$/$\Delta\lambda\sim$300.
Exposures of the dome interior illuminated with incandescent lamps
through the broadband K filter were used for flat-fielding.
A set of observations consisted of the star
observed at two positions along the slit, separated by 10 arcsec.
A typical exposure time for the brown dwarf candidates
was 120 seconds per slit position
and limited by variations
in the atmospheric OH emission.  Total on-source integration times
were typically 2 minutes for the standard star observations 
and ranged from 8 to 36 minutes for the brown dwarf
candidates (see Table 2).

All data were reduced using IRAF.
Before extraction, the spectral images in each set were sky-subtracted
using their companion image,
and divided by a normalized, dark-subtracted flat-field.  Wavelength calibration was established
using OH emission lines.  After extraction, the spectra were median-combined after
matching their wavelengths.
The resulting spectra were
shifted to the same wavelength scale as a
a spectrum of either the A0V star HR~1237 or the A1V star HR~1027
that was observed close in time and airmass by cross-correlating telluric absorption lines.
Telluric features
in the combined spectra were removed by dividing by the appropriate A star spectrum.
No attempt was made to restore the true continuum shape of
the program objects; to recover the true shape
one could multiply each spectrum by a blackbody spectrum
corresponding to the effective temperature of the telluric standard.
Because of the steep response function of the grism, it was not
possible to extrapolate over the Br $\gamma$ absorption line in the
telluric standard and hence we were not sensitive to
Br $\gamma$ emission in the program stars.
The signal-to-noise ratios achieved in the normalized
continua (SNR$_{cont}$) of the reduced spectra are
listed in the last column of Table 2.

\section{Spectroscopic Analysis}

\subsection{Spectral Classification Using a Water Vapor Index}

In the K-band, water vapor absorption bands centered at 1.9 $\mu$m and 2.5 $\mu$m
provide the best means for deriving spectral types for reddened M stars.  These bands are not
only sensitive to temperature for cool stars (Jones \etal 1994) but also are well-resolved
in low resolution spectra.  Water vapor indices have been used to derive spectral
types for very low mass YSOs in $\rho$ Ophiuchi (WGM99, Cushing, Tokunaga, \& Kobayashi 2000), 
IC~348 (Najita, Teide, \& Carr
2000), the ONC (Lucas \etal 2001), the Chamaeleon cloud (Gomez \& Persi 2002),
and the Taurus cloud (Itoh, Tamura, \& Tokunaga
2002).  
In WGM99, we defined a Q index that was independent of reddening as a measure of the depth
of the water vapor absorption:
\begin{equation}
Q = (F1/F2)(F3/F2)^{1.22}
\end{equation}
where F1, F2, and F3 are the average relative fluxes in narrow bands
covering 2.07--2.13 $\mu$m, 2.2695--2.2875 $\mu$m, and 2.40--2.50 $\mu$m, 
respectively.
The uncertainty in Q was calculated by using Eqn. 2 and propagating error estimates for the mean
fluxes in F1, F2, and F3.  The error in F$_i$ is given by $\sigma_{F_i}$ = F$_i$/($\sqrt{N} 
\times SNR_{cont}$) 
where N is the number of channels in F$_i$ and $SNR_{cont}$ is the signal-to-noise ratio
in the continuum.  This assumes the flux uncertainty in each channel of the band F$_i$ is the same.
As a check, we calculated the uncertainty in the Q values for 3 
brown dwarf candidates by calculating the standard deviation of the ensemble
of Q estimates from
individual spectra.  The two estimates are in excellent agreement.\footnote
{Uncertainties in Q quoted in WGM99 used the sample standard deviations of the average
values in the bands F1, F2, F3 and overestimate the true error in Q.}

A linear fit to a plot of the Q index vs. the optically-determined
M spectral type for dwarf standards weighted by the uncertainties in Q
for stars from M0.5V--M9V
yields the following relation
\begin{equation}
MV~subclass = (-20.20\pm0.30) \times Q + (18.41\pm0.35)
\end{equation}
with a correlation coefficient of r=0.98 for a sample of 12 measurements.
This relation supercedes that in WGM99 as it utilizes an improved wavelength calibration, 
a redefinition of the F2 band, and a better estimate of the uncertainty in Q.
This relation is plotted in blue in Fig. 3. 

\subsection{The Water Index as a Function of Surface Gravity}

A sample of seven subgiants were selected for infrared spectroscopic
observations from the MBM~12 and IC~348 YSO
populations (Hearty \etal 2000; Luhman 2001; Luhman \etal 1998; Luhman 1999).
The median ages of these clusters are estimated to
be 2 Myr and the median ages of the sources in our sample is 1 Myr.  
As shown in Fig. 3 in green, the linear fit to the subgiant sample is
consistent with the dwarf relation and there is no strong sensitivity of the
Q index to surface gravities in the dwarf to subgiant range.  
This is in agreement with the findings of Gorlova \etal (2003) who found the linear
relationship between the water index and M spectral type of young cluster objects
(log(g)=3.68$\pm$0.42) closely resembled that of a sample of field stars (log(g)=5.17$\pm$0.38).

The Q values for the 8 giants observed show a much greater scatter 
from a linear fit (shown in red in Fig. 3).  
In our sample, the giant star spectra are distinct from those of dwarfs
and always display weaker water vapor absorption than dwarfs
for a given spectral type (see Appendix A).  
band is not evident until a spectral type of M7III.   
Yet, the Q index for 
giants can be smaller than that of dwarfs of the same spectral type 
due to enhanced absorption by CO that 
mimics water vapor absorption in the 
2.40--2.50 $\mu$m band.
Variability must also
contribute to the scatter in the giant sample; all of the stars from M4-M9
are known variables and two (VY~Peg and AW~Psc) are
Mira variables.  We conclude that the Q index is not a useful measure of spectral
type for objects with the surface gravities of giant stars.

\section{Properties of the Brown Dwarf Candidates}

\subsection{Spectral Types}

We observed one-half of the 50 sources in an extinction-limited sample defined in Fig. 2 
with estimated masses of M=0.04--0.10 M$_{\sun}$ 
and A$_v$ $<$10 mag.
All 25 brown dwarf candidates displayed late-type photospheres with absorption due to 
water vapor and CO.  Spectra are presented in Fig. 4.  
Since the surface gravities
of objects in this sample are likely to resemble closely those of the subgiants observed,
spectral types were derived using Eqn. 3 which is appropriate for both subgiants
and dwarfs.
Q values and the resulting spectral classifications for the brown dwarf candidates are 
presented in Table 3. 
Uncertainties in the Q index were $\pm$0.010--0.022 due to statistical errors alone leading
to uncertainties in the M spectral type of $\pm$0.2--$\pm$0.4 subclasses.
The derived spectral types vary between M2.5 and M8.0, with 17 of 25 
candidates displaying spectral types between M6.0 and M8.0.  
Visual extinctions were derived using the relation A$_v$ = 9.09$\times$[$(J-H)_{obs}$ - $(J-H)_
0$] where $(J-H)_0$ is the instrinsic color for a given spectral type (WGM99).  
The values for A$_v$ are 
$\le$4 mag for all but 2 sources
and $\le$7 mag for all sources, consistent with their locations in the 
color-color diagram (Fig. 5).

The majority of the candidates appear to be YSOs of very low mass.  The projection of
this sample on the high column density core (A$_v$ $>$ 10 mag), coupled with the relatively
low visual extinctions of the candidates, minimizes the chance that some objects are
background M stars.  In addition, fifteen of the 25 candidates are known x-ray sources, including
3 tentative detections (Getman \etal 2002, Preibisch 2003).  The detection of x-ray emission 
argues in favor of the youth of this sample, as x-ray luminosity  
and stellar activity are
known to decline with age for K and M stars (Fleming, Schmitt, \& Giampapa 1995).  
Finally, we estimate the number of foreground stars at this galactic latitude and longitude
using star count models.  Using the Wainscoat \etal (1992) models, which have been found to agree
with the observed star counts from the 2MASS survey to within 15\% (Carpenter 2000), we predict
at most 2 foreground objects in the sample of 25.

\subsection{Infrared Excesses}

There are no strong infrared excesses among the spectroscopic
sample of brown dwarf candidates.
The candidates are plotted on a (J-H) vs. (H-K) diagram in Fig. 5 relative to the loci
of giants, dwarfs (through M6), 
and classical T Tauri stars (CTTS, Meyer, Calvet, \& Hillenbrand 1997).  
Objects that lie to the right of the reddening band for main sequence stars
have possible K excesses.  ASR~79 lies to the left of the reddening band, a likely
result of an associated infrared nebula known optically as HH~4.
In addition, we have computed the excess emission at K, defined as r$_k$=F$_{K_{ex}}$/F$_{K_*}$, 
for each source where F$_{K_{ex}}$ is the K-band flux from circumstellar emission and
F$_{K_*}$ is the expected stellar flux at $\lambda$=2.2 $\mu$m.\footnote
{r$_k$ is calculated by comparing dereddened observed colors to intrinsic colors, i.e., r$_k$ = 
F$_{Kex}$/F$_K$ = [(1+r$_h$)(10$^{[(H-K)-(H-K)_0-0.065*A_v]/2.5}$)]-1.  Our estimates for r$_k$
are lower limits since, by dereddening to intrinsic dwarf colors, we have assumed r$_j$=r$_h$=0 
(e.g., Meyer \etal 1997).}
These values are presented in
Table 3.
While moderate K-band excesses are indicated
for ASR~38 and MBO~79, the majority of objects have colors 
consistent with no K-band excess.
As noted in WGM99, strong veiling by dust can lead one to derive
spectral types using the Q index that are too early by 2-3 subclasses.  It is
possible that the calculated spectral types for ASR~38 and MBO~79 are too early by
1--2 subclasses due to moderate veiling, but it is expected that 
the spectral types for the remaining 
objects are relatively unaffected.

Objects without strong K-band excess emission may also
possess circumstellar disks.  Disk luminosities are expected to be lower for very low
mass stars and 
substellar objects resulting in cooler inner disks and little or no K-band excess.
Such a trend has been observed in the ONC (Hillenbrand \etal 1998).
Indeed, L-band surveys have been shown to be more sensitive to the 
thermal disk emission in low mass objects (Haisch, Lada, \& Lada 2001).  A recent L\amin\
survey by Liu, Najita, \& Tokunaga (2003) has revealed that about 80\% of the young
brown dwarfs in their sample showed evidence for circumstellar disks.

\subsection{Masses and Ages of the Brown Dwarf Candidates}

We followed the procedure outlined in WGM99 to 
estimate the masses and ages of the brown dwarf candidates.
Briefly, the adopted temperature scale for M dwarfs derived in WGM99 is based on the modified
blackbody fits of Jones \etal (1996).  This temperature scale overlaps with the scale adopted
by Dahn \etal (2002) for M5--L8 dwarfs and is consistent
within mutual uncertainties.
Effective temperatures for each brown dwarf candidate derived in this manner are
given in Table 3.  Their typical
uncertainties of $\pm$95 K or $\pm$0.016 dex are due to nearly equal contributions from the
errors in the determination of the spectral type and the temperature scale. 
However, systematic effects could affect the derived effective temperatures.
As alluded to earlier, the presence of moderate veiling in 2 objects could
lead us to overestimate their effective temperatures by $\sim$150 K.
The assumption of dwarf, rather than subgiant, surface gravities could
also systematically affect our results.
For spectral types of M2 and later, giant star temperatures are
warmer than dwarfs by 300--500 K for stars of the same spectral type 
(for example see Fig. 4 in Itoh \etal 2002).  Hence by assuming a dwarf surface gravity, 
we may be underestimating the effective temperature of an M2 subgiant by $\sim$150 K 
and of an M6--M9 subgiant by $\sim$250 K.    
With these caveats in mind, we adopt the temperatures listed in Table
3 for our brown dwarf candidates assuming a dwarf temperature scale.
Values for Log(L$_{bol}$/L$_{\sun}$) were derived from the absolute J magnitude and 
are presented in Table 3.  The formal uncertainties of Log(L$_{bol}$/L$_{\sun}$) are
$\pm$0.18 dex and were estimated by propagating errors in the bolometric correction, the J
magnitude, the distance modulus, and the extinction at J.

In Fig. 6, we have plotted our 25 brown dwarf candidates on Hertzsprung-Russell diagrams 
overlaid with the theoretical tracks and isochrones from the models of
of D'Antona \& Mazzitelli (1997,1998, hereafter DM), Baraffe \etal (1998, hereafter BCAH), 
and Burrows et al. (1997, 1998).  The DM models use opacities from Alexander \& Ferguson (1994), the full
spectrum of turbulence convection model of Canuto \& Mazzitelli (1991), and assume mass
fractions of helium, metals, and deuterium of Y=0.28, Z=0.02, and X$_D$=2 $\times$ 10$^{-5}$.
BCAH models have been extended to include objects with masses as low as 0.02 M$_{\sun}$ 
and assume a mass fraction
of helium Y=0.275, solar metallicity, and a general mixing length parameter of
$\alpha$=1.
The Burrows et al. models are constructed specifically for giant planets and brown dwarfs
and assume mass fractions of helium, metals, and deuterium of Y=0.25, Z=0.02,
and X$_D$=2 $\times$ 10$^{-5}$.
As found in most young clusters, there is a scatter in age due in part to real age differences
between sources and to uncertainties in luminosities that arise from variations in 
distance and errors introduced by the photometry and dereddening 
(the latter are represented by the error bar in Fig. 6).  Ages can be overestimated
due to underestimating the luminosity of a source or overestimating its temperature.
The former could be due to underestimating the extinction toward the source owing to 
unresolved scattered light (e.g., ASR~79) 
and the latter could be due to an infrared excess as discussed in Sec. 4.2
(e.g., MBO~79). 

For the DM models (Fig. 6a), the brown dwarf candidates have masses ranging 
from $<$0.02 M$_{\sun}$ to 0.25 M$_{\sun}$, with 16 objects at or
below the hydrogen-burning limit.  The probable brown dwarfs are indicated in the
last column of Table 3.  The median age for the sample implied by the models is 0.3 Myr.
Similar ranges in ages and masses are implied by considering the Baraffe et al models (Fig. 6b).
The Burrows et al. models generally yield lower masses and younger ages with 23 objects at or
below the hydrogen-burning limit.  The median age for the sample implied by these models is
$<$0.3 Myr.  

\subsection{Surface Gravity Estimates}

Surface gravities (g=GM/R$^2$) have been estimated for the brown dwarf
candidates using the DM mass estimates and the radii
derived from the luminosity and effective temperature.
Values range from log(g)=3.0 to 4.6 in cgs units with a median value
of 3.3.  These estimates assume a temperature scale, intrinsic colors,
and bolometric corrections derived from dwarf standards.
For comparison, the log(g) of the Sun is 4.4 (Livingston 1999).
Surface gravity
estimates {\it increase} under the assumption of giant properties, primarily due
to their higher effective temperatures for a given M spectral type, implying that our estimates
of log(g) are lower limits.  
Thus, as shown in Fig. 7, the surface gravities of our objects
more closely resemble dwarfs than giants, implying that our use of a dwarf temperature
scale has led to an underestimation of the temperature for M6--M9 stars of no more
than $\sim$250 K.

The dominant source of uncertainty in our mass estimates is the surface gravity
dependence of our temperature determinations.  The arrows in Fig. 6
indicate the maximum shift in temperature for an object of a given spectral type
if its temperature is midway
between that of a dwarf and giant rather than that of a dwarf as assumed.  While as
many as 16 of our candidates have masses at or below the hydrogen burning limit (in the DM models),
only 9 are cool enough to be classified as brown dwarfs after undergoing this shift.
In Table 3 we distinguish between objects that lie 250 K or more to the right of the hydrogen-burning
limit (``bd" in last column) and those to the right of the limit but within 250 K (``bd?").
In the group of nine brown dwarfs is the moderately-veiled source ASR~38,
and 4 x-ray sources including the x-ray flare source ASR~24.


\section{Properties of the Northern Cluster}

To explore further the properties of the young stellar and substellar population in NGC~1333,
we have defined a subsample
of 141 sources from our JHK survey of the northern cluster that lie within 
a 79 arcmin$^2$ box
enclosing the highest column density gas.  The southwest corner of 
the box is located at RA(J2000)=
3$^h$ 28$^m$ 53$^s$, DEC(J2000)=+31\deg\ 18\amin\ 19\asec\ and the
northeast corner at  RA(J2000) = 3$^h$ 29$^m$ 30$^s$, DEC(J2000)=
+31\deg\ 28\amin\ 19\asec.
The average visual extinction (as 
implied by the C$^{18}$O 
column densities) is greater than 10 mag over 85\% of the area.
The total mass of gas from these data
is estimated to be 100 M$_{\odot}$, with a factor of 2 uncertainty (Hatchell \etal 2003).

\subsection{Membership and Stellar Density}

The high column density gas in this region should minimize the contamination 
of the sample by background field stars. 
To estimate the degree of contamination, 
we used the C$^{18}$O column densities to subdivide
the cloud core into 6 extinction regions as defined by the contours in Fig. 1 with
average visual 
extinctions of 6.5, 13.5, 20.5, 27.5, 34.5, and 41.5 mag. 
Galactic star count models 
(Wainscoat \etal 1992)
were then used with
the corresponding extinction screen to predict the number of foreground and
background stars in the field.  We find that
of the 141 infrared 
sources in this region, it is estimated no more than 15 are field stars.
For an effective radius of 0.44 pc ( $R_{eff}$ = $\sqrt{A/\pi}$ where A is the area of the box),
this suggests a stellar density of $\sim$360 stars pc$^{-3}$.
Such stellar densities 
are typical of infrared clusters and underscore their stability
against tidal disruption by nearby clouds while they are gravitationally
bound by the molecular gas (e.g., Wilking 2001).

\subsection{An Extinction-Limited Sample}

Using the median age derived for the spectroscopic sample, we will define an unbiassed
sample of YSOs to explore the disk frequency and mass function of the northern cluster.
First, estimates for the extinction toward
each member of subsample are needed.
Depending on their location in the (J-H) vs. (H-K) color--color diagram,
and the M$_J$ vs. (J-H) color--magnitude diagram, objects
were dereddened either to the CTTS locus (color--color
diagram) or to a pre-main sequence isochrone (color--magnitude diagram).  
If the (J-H) vs. (H-K) colors of objects in our sample lie within
the reddened CTTS locus (Fig. 5), {\it and} the dereddened absolute
J--band
magnitude is consistent with a mass greater than 0.1 M$_{\odot}$,
the source was dereddened to the CTTS locus.  The CTTS locus
is only valid for stellar mass objects with spectral types earlier than
M6 which corresponds roughly to the temperature of an
object near the hydrogen--burning limit for the age of the cluster.
For objects that
lie within the reddened main sequence in the color--color diagram, as
well as those within the CTTS locus whose dereddened M$_J$ suggest
a mass below 0.1 M$_{\odot}$, the reddening was estimated by projecting
objects
in the color--magnitude diagram to an
isochrone consistent with the cluster age derived above from the
H--R diagram.

Once the extinction has been estimated, then
the absolute J--band magnitude can be calculated.
Using intrinsic colors and bolometric corrections for
very late--type dwarf stars (see Appendix A in WGM99), we transformed
the DM models to estimate the mass from the absolute J magnitude.
In order to define a
complete sample unbiassed toward stars of higher mass,
we defined an {\it extinction--limited sample} over a fixed
mass range.  
Adopting an age of 0.3 Myr for the low mass cluster members as suggested by
the spectroscopic observations,
we sampled 46 objects down to 0.04 M$_{\odot}$ with A$_V <$ 13$^m$.
This is the minimum mass for which we can reliably estimate
an age from our spectroscopic survey.
Of these, two objects exhibit extreme infrared colors similar
to flat--spectrum and Class I sources in the Taurus dark cloud
(cf. Kenyon and Hartmann, 1995).
These objects cannot be dereddened and are removed from the sample.
Of the remaining 44, five exhibit colors similar to early--type (B--A--F)
YSOs (e.g. Hillenbrand et al. 1992; Lada and Adams,
1992), 12 are
stellar mass objects within
the reddened main sequence,
11 are stellar mass objects within the reddened CTTS locus, 
and 16 are brown dwarf candidates.

The total mass of stellar and substellar objects in this sample is about 8 M$_{\odot}$.  
For the 7 objects in common between the extinction-limited sample and the spectroscopic
sample, mass estimates for individual objects agree to within a factor of 2 which is
consistent with previous applications of this technique (Meyer \etal 2000).  It is
important to note, however, that the mean mass of the two samples is virtually identical (within 10\%)
emphasizing that the assumption of a common age yields a meaningful mass estimate
for a statistically significant ensemble.
The mass of gas associated this region where A$_V <$ 13$^m$ is estimated to be $\sim$70 M$_{\odot}$.  
Given the total mass of young stars, the efficiency of star
formation would be 10\%, with a factor of 2 uncertainty from the mass estimate alone.   This value is 
consistent with that found in other young clusters and suggests the eventual dispersal of the
cluster once the gas is removed (Carpenter 2000; Lada \& Lada 2003).
 
\subsubsection{Disk Frequency}

The fraction of objects
exhibiting IR excesses consistent with the presence of active accretion
disks within 0.1 AU of the star can be determined from an object's location in the color-color diagram (Fig. 5).
Within the A$_V$--limited sample of {\it stellar} mass objects defined here ($>$ 0.1
M$_{\odot}$), the fraction of YSOs displaying an IR excess
is $F_{IRX} = 0.57 \pm 0.13$ adopting 
Poisson errors.  
Because some T Tauri stars have large
inner holes in their disks that removes hot dust associated with
JHK excess emission, this inner disk fraction can underestimate
the fraction of stars with active accretion disks.  Again, using
the Taurus dark cloud as a guide (Meyer \etal 1997), we estimate the corrected fraction of
active accretion disks at $F_{acc} = 0.75 \pm .20$.
These estimates are in excellent agreement with those determined for the entire NGC
1333 cluster from the SQIID survey by LAL and consistent with disk
frequencies estimated for clusters of comparable age (e.g., Haisch \etal 2001).
We are not able to assess the disk
frequency for the substellar sample using the JHK
color--color diagram alone because the intrinsic colors
of late type stars overlap those expected for active
accretion disks.  Complete spectroscopic samples or longer
wavelength photometry are required to assess the disk
frequency of ensemble of the substellar objects in NGC 1333.

\subsubsection{Mass Spectrum}

While our sample is not large enough to examine the detailed shape of the cluster's
mass spectrum, we can explore its general properties.
Among stellar objects, we can calculate the ratio of
intermediate mass (1-10 M$_{\odot}$) to low mass (0.1-1 M$_{\odot}$) stars.
The ratio ${\cal R} = 0.08 \pm 0.04$ with the error computed assuming Poisson statistics.
This ratio is consistent with being drawn from the field star IMF (e.g., Miller \&
Scalo 1979; Kroupa, Tout, \& Gilmore 1993) and is similar to $\cal R$ values
computed from other clusters such as the ONC, Mon R2, Rho Ophiuchi,
and IC 348 (Meyer \etal 2000).

Turning to the subset of cluster members whose M$_J$ suggests a
mass below 0.1 M$_{\odot}$, we determine the ratio of stellar
(0.1-1 M$_{\odot}$) to substellar mass (0.04- 0.1 M$_{\odot}$) objects 
within our extinction--limited sample.  The ratio ${\cal R}_{ss} = 1.1 +0.8/-0.4$
which is a lower limit given possible contamination of the
substellar sample by field stars (Sec. 5.1).   
Hence our value of ${\cal R}_{ss}$ sets an upper limit to the slope 
for the low mass end of
the mass spectrum of $\alpha \le$1.6 where dN/dM $\propto$ M$^{-\alpha}$.
This suggests that the mass of the population in the mass range of 0.04-1 M$_{\odot}$
is dominated by low mass stars
and not brown dwarfs.  This upper limit is consistent with values of $\alpha$=1--2
derived from the
observed distribution of L dwarfs in the solar neighborhood (Reid \etal 1999).  Our
upper limit is also consistent with the slopes derived for other young clusters;
these estimates range from $\alpha$=0.43 for the ONC (Hillenbrand \& Carpenter 2000) and
$\alpha$=0.5 for IC 348 and $\rho$ Oph (Najita \etal 2000; Luhman \& Rieke 1999) to
$\alpha$=0.8 to 1.2 for $\sigma$ Orionis cluster (B\'ejar \etal 2001; Tej \etal 2002).

\subsubsection{Sensitivity to Age}

If we instead take an age of 1 Myr for the cluster associated with
NGC 1333, perhaps more appropriate for the mass range 0.1--2.5
M$_{\odot}$ (Aspin 2003; Luhman, priv. comm.),
our results do not change significantly.  A trend of increasing age with
stellar mass is observed toward many young clusters (e.g., Hillenbrand 1997)
and it is suspected that this is
a problem with the isochrones rather than evidence that the star formation
rate varies with age {\it and} mass (Palla \& Stahler 1999).
For an age of 1 Myr, our extinction--limited sample becomes
52 objects with A$_v < 10^m$.  Of these, two have extreme
colors similar to protostars and cannot be dereddened as described
above.  Of the 50 remaining objects in the sample, five have IR excesses
similar to ``early--type'' stars, 14 are within the stellar reddened main sequence,
11 lie within the CTTS
locus,
and 20 are brown dwarf candidates.  
The IR excess fraction is virtually unchanged with $F_{IRX} = 0.59 \pm 0.15$ and with a
corrected active accretion disk fraction of $F_{IRX} = 0.80 \pm 0.20$.
The ratio of intermediate
to low mass stars is ${\cal R} = 0.18 \pm 0.14$ which is also 
consistent with being drawn from the field star IMF.  This difference,
although formally in agreement with the results obtained assuming an
age of 0.3 Myr, is to be expected
given the luminosity evolution of pre-main-sequence stars over time.
Finally, the 
ratio of stellar to substellar mass objects is ${\cal R}_{ss} = 0.6 \pm 0.2$
which implies an upper limit to the slope for the mass spectrum at low masses of $\alpha\le$2.
While formally consistent with the results quoted above, this
highlights the sensitivity of our estimate of the substellar
IMF with age.

\section{Summary}

We have conducted a new near-infrared survey of the northern cluster embedded in the NGC 1333
cloud.  When combined with previous surveys, near-infrared magnitudes and colors were
used to select 25 brown dwarf candidates in the high-extinction cloud core.
Low resolution infrared spectroscopy in the K band revealed cool photospheres for all
of the candidates.  Spectral classifications were achieved using a water vapor index
that is independent of reddening and insensitive to surface gravity in the range expected for
subgiants and dwarfs.  The derived spectral types varied between M2.3 and M8.2, with 17 of 25
candidates displaying spectral types between M6.0 and M8.2.  Based upon their superposition on the
high extinction cloud core, their high incidence of x-ray activity, and the low number of foreground
objects predicted by star count models, it is expected that at least 90\% of the sample are 
very low mass members of
the NGC 1333 cluster.

Once spectral types and effective temperatures were determined, 
stellar luminosities were estimated.  Comparisons of these properties with three sets
of theoretical models suggest a median age for the sample of $<$1 Myr
and a substellar mass for at least 9 of the candidates including the x-ray flare source ASR~24.  
Another 7 candidates lie near
the hydrogen-burning limit and are possible brown dwarfs.  The largest uncertainty
in our mass estimates arises from the effective temperature and the sensitivity of the temperature scale
to surface gravity.  We have shown that surface gravities estimated for the brown 
dwarf candidates more closely
resemble dwarfs than giants for a given spectral type and hence a dwarf temperature scale
is more appropriate than a giant temperature scale for young objects.

Finally, we have used our near-infrared survey to define an extinction-limited sample of
YSOs in the high-extinction core to investigate the properties of the
northern cluster.  For this sample of 44 objects, the
the fraction of objects exhibiting IR excesses is $F_{IRX} = 0.57 \pm 0.13$, in
excellent agreement with previous estimates for this cluster and other clusters
of comparable age.  Using masses estimated from intrinsic colors and absolute J magnitudes for
very late--type dwarf stars and the 0.3 Myr DM models, we find a mass spectrum
for intermediate and low mass stars that is indistinguishable from that in other young clusters and
the field star IMF.  Due to the possible presence of background stars in the sample, only
an upper limit to the slope of the mass spectrum at very low and substellar masses can
be estimated.  The slope of $\alpha \le$1.6 is consistent with that found 
for substellar objects in other young clusters and in the field.

\acknowledgments

We would like to thank Marcia Rieke, George Rieke, and Erick Young for 
their assistance with the infrared survey at Mt. Bigelow Observatory.
Marcel Bechtoldt, Tony Denault, and Alan
Tokunaga are gratefully acknowledged for their efforts in establishing the Internet2
connection from St. Louis to Mauna Kea for the IRTF observations.
We would also like to thank Gary Fuller, Jennifer Hatchell, Charles Lada, Kevin Luhman,
and Konstantin
Getman for their generosity in sharing data in advance of publication.  
We are grateful to Mike Merrill for his assistance with SQMOS
and to Margaret Meixner and Rich Schuler for their efforts with a preliminary NGC 1333 survey using
NIRIM at Mt. Laguna Observatory.  We benefited from many discussions with John Carpenter,
Nadja Gorlova, and Lynne Hillenbrand.  BW and AM gratefully
acknowledge support from RUI Grant NSF AST 98-20898 and the NASA/Missouri Space Grant Consortium
and GC from the Missouri Research Board.

\appendix
\section{Spectra of M Giants}
Grism spectra were obtained for eight objects of low surface gravity and are presented in Fig. 8.
Owing to the high infrared brightness of these objects, the SNRs in the unsmoothed spectra
are greater than 100.
The giant sample includes
the following objects and their optically-determined
spectral types: HD~236879 and BD+30~3647 (M0, Jaschek \etal 1978),
BD+29~3674 and BD+06~4112 (M2, Jaschek \etal 1978), HD~167654 (M4, Houk \& Swift 1999), BD+27~3478
(M5, Jaschek \etal 1978), VY Peg (M7, Kirkpatrick \etal 1997), CTI+191258.9+280353
(M7, Kirkpatrick \etal 1994), and AW~Psc (M9, Solf 1978).  We note that the luminosity class
for AW~Psc is not well-determined.

Features in the giant spectra are distinct from those in the cluster member or
dwarf star spectra.  The giant star
spectra are dominated by the CO bandheads and display weak Na I and Ca I absorption lines.
In addition to the four $^{12}$C$^{16}$O  ($\delta\nu$=2) bandheads normally seen in infrared spectra,
also seen is the (6,4) bandhead at 2.414 $\mu$m, the (7,5) bandhead at 2.446 $\mu$m, 
and perhaps even 
the (8,6) bandhead at 2.479 $\mu$m (see Goorvitch 1994 for line list).  This leads to an overall
depression of the continuum in the 2.4--2.5 $\mu$m region and hence a Q index such as 
defined in Eqn. 2 
cannot be used to measure the depths of the water bands.
Water vapor absorption at 1.9 $\mu$m is first evident in the spectrum of the M7~III VY~Peg but is 
weak or absent in the  M7~III VY~Peg CTI+191258.9+280353.  This is no doubt related
to the pulsation of these variable stars which leads to fluctuations in their effective temperatures.

\pagebreak

\begin{center}
{\bf Figure Captions}
\end{center}
\medskip
%
%
\figcaption{%
The distribution of the 25 brown dwarf candidates relative to the high column density gas
in the NGC 1333 core.  The boxes outline the extents of deep infrared surveys 
of the northern cluster (this group)
and the southern cluster (Aspin \etal 1994).  Coordinates are in J2000.0.
Contours correspond to the integrated intensity
of C$^{18}$(1-0) emission in units of T$_{MB}$ beginning at 2 K km s$^{-1}$ and increasing in steps
of 0.7 K km s$^{-1}$.  The contours correspond roughly to A$_v$ values of 10, 17, 24, 31, 38, 45, and 52 mag.
The C$^{18}$(1-0) data were obtained at FCRAO and smoothed to a resolution of 1 arcmin (Hatchell \etal 2003,
in preparation).
}
%
\figcaption{%
A color-magnitude diagram for the NGC 1333 double cluster.
Photometry
is taken from this study (northern cluster) or from the survey of Aspin \etal (1994) (southern cluster), 
transformed into the CIT system.
The Aspin \etal (1994) JHK photometry
for the southern cluster was transformed into the
CIT system using the formulation of Casali \& Harwarden (1992).
Isochrones are shown for 1 Myr (0.02--2.0 M$_{\sun}$) 
and for 5 $\times$ 10$^8$ years (0.1--2.0 M$_{\sun}$, labeled ``ZAMS").
Reddening lines from the 1 Myr isochrone are
shown for selected masses by dashed lines and were calculated using the
extinction law derived by Cohen \etal (1981).  For comparison, a reddening
vector for an A$_v$=10 mag is also shown.
}
%
\figcaption{%
The Q index vs. M spectral type for dwarf standards (blue), subgiants (green), and
giants (red).  Linear fits to the dwarf, subgiant, and giant data are shown in the
same color scheme.  The dwarf sample is from WGM99.
The subgiants in MBM~12 and their optically-determined 
spectral types include:
LkH$\alpha$262 (M0), LkH$\alpha$263 (M3), S18 (M3), and RXJ0258 (M4.5).
The subgiants in IC~348 and their optically-determined spectral types include:
Luhman source IDs 158 (M5), 312 (M6), and 355 (M8).  The giant sample is described in Appendix A.
}
%
\figcaption{%
Infrared spectra for the 25 brown dwarf candidates observed with the IRTF grism.
Spectra have been smoothed to a resolution
of R=170 for display purposes.  The removal of telluric absorption lines, accomplished by dividing
a spectrum by that of an AV star, leaves a false emission line at the wavelength of
B$\gamma$ which has been blanked out.  The wavelengths of absorption lines of Na, Ca, and 
6 CO bandheads are marked at the bottom of the plot.  The three wavelength ranges averaged
to compute the water vapor index Q
are labeled at the top of the plot as F1, F2, and F3.  Fig. 4a shows spectra for the sources
with spectral types ranging from M2.3 (top) to M5.6 (bottom) as indicated by the increasing
depth of the water vapor bands. 
Fig. 4b shows spectra for the sources ranging from M6.2 to M7.4 and Fig. 4c for sources
ranging from M7.4 to M8.2.
}
%
\figcaption{%
A $(J-H) vs. (H-K)$ diagram for the brown dwarf candidates.  Photometry is from
this study, the study of Aspin \etal (1994), or from Lada \etal (1996) transformed into the CIT
photometric system.  The loci for the intrinsic colors of dwarfs and giants
were adopted from Bessell \& Brett (1998), using their conversion to the CIT
system.  The loci for classical T Tauri stars is adopted from Meyer \etal (1997).
Object symbols for identified x-ray emitters are filled with an ``X".  An error bar of
$\pm$0.07 mag is shown that would arise in the colors from uncorrected variations in the array
responsivity due to vignetting.
}
%
\figcaption{%
Hertzsprung-Russell diagrams for the NGC~1333 brown dwarf candidates assuming a distance of 300 pc.
X-ray emitters are denoted by an ``X" and non-xray emitters by a solid diamond. 
The positions of the candidates are shown relative to the theoretical tracks and isochrones
of D'Antona \& Mazzitelli (1994,1997,1998) in Fig. 6a, relative to the
models of Baraffe et al. (1998) in Fig. 6b, and relative to the models of Burrows \etal (1997,1998)
in Fig. 6c.  Isochrones from
10$^5$ years to 10$^8$ years are shown by solid lines and evolutionary tracks
from 0.01 M$_{\sun}$ to 0.40 M$_{\sun}$ are shown by dashed lines.
The bold dashed line marks the evolutionary track for a star at the
hydrogen-burning limit.  
The typical error bar
for a candidate is shown in the lower left of each plot and is
$\pm$0.014 dex in Log(T$_{eff}$) due to uncertainties introduced by the calculation of
the Q index and the fit to derive the dwarf temperature scale.  The error of $\pm$0.18 dex in
Log(L$_{bol}$/L$_{\sun}$) is dominated by uncertainties in the distance ($\pm$30 pc) and in the
extinction at J ($\pm$0.37).  An arrow labeled ``250 K"
indicates the maximum systematic error introduced by assuming a dwarf temperature scale for a
subgiant star.
}
%
\figcaption{%
The surface gravities in units of cm s$^{-2}$ for the brown dwarf candidates estimated using their
masses using the DM tracks.  The YSO values are shown by star symbols.
For comparison, the surface gravities for
dwarf (open triangles) and giant stars (open squares) are from the compilation of
Drilling \& Landolt (1999).  
}
%
\figcaption{%
Spectra for the M giants observed with the IRTF grism.  Spectra have been smoothed to a resolution
of R=170 for display purposes.  The removal of telluric absorption lines, accomplished by dividing
a spectrum by that of an AV star, leaves a false emission line at the wavelength of
B$\gamma$ which has been blanked out.  The wavelengths of absorption lines of Na, Ca, and 
6 CO bandheads are marked at the bottom of the plot.  The three wavelength ranges averaged
to compute the water vapor index Q
are labeled at the top of the plot as F1, F2, and F3.}

\newpage
\centerline{\includegraphics{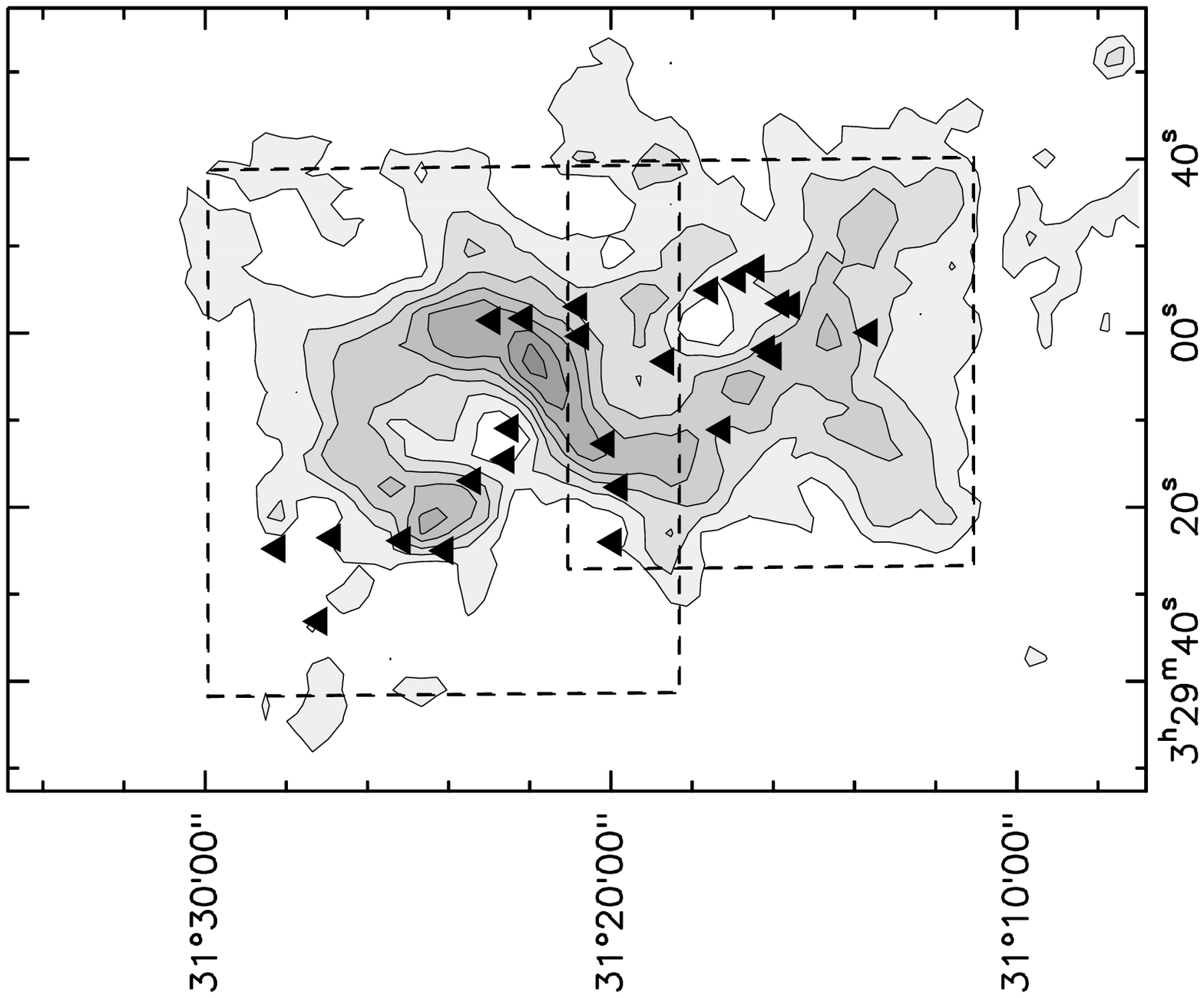}}
\newpage
\centerline{\includegraphics{Wilking.fig2.ps}}
\newpage
\centerline{\includegraphics{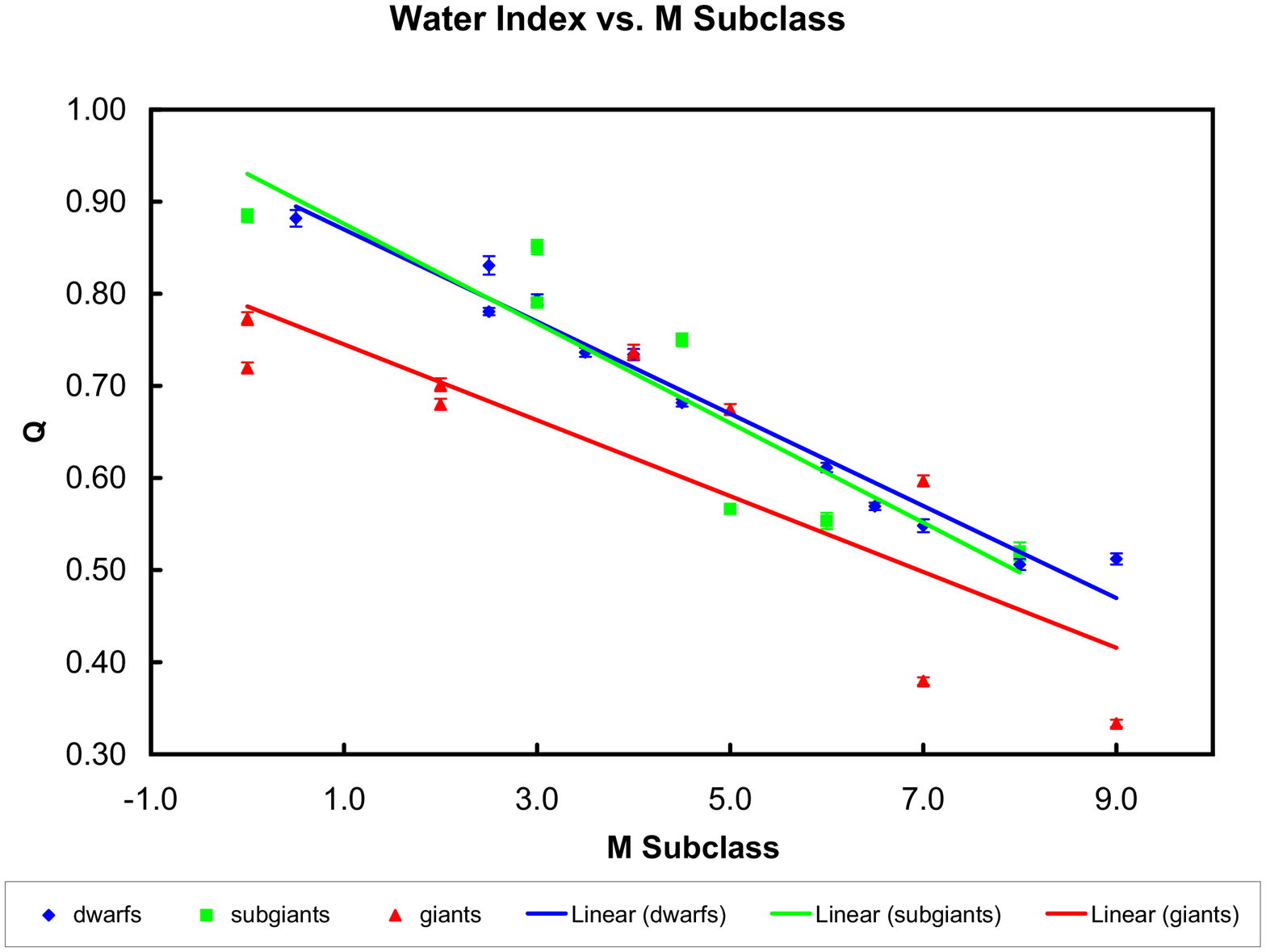}}
\newpage
\centerline{\includegraphics{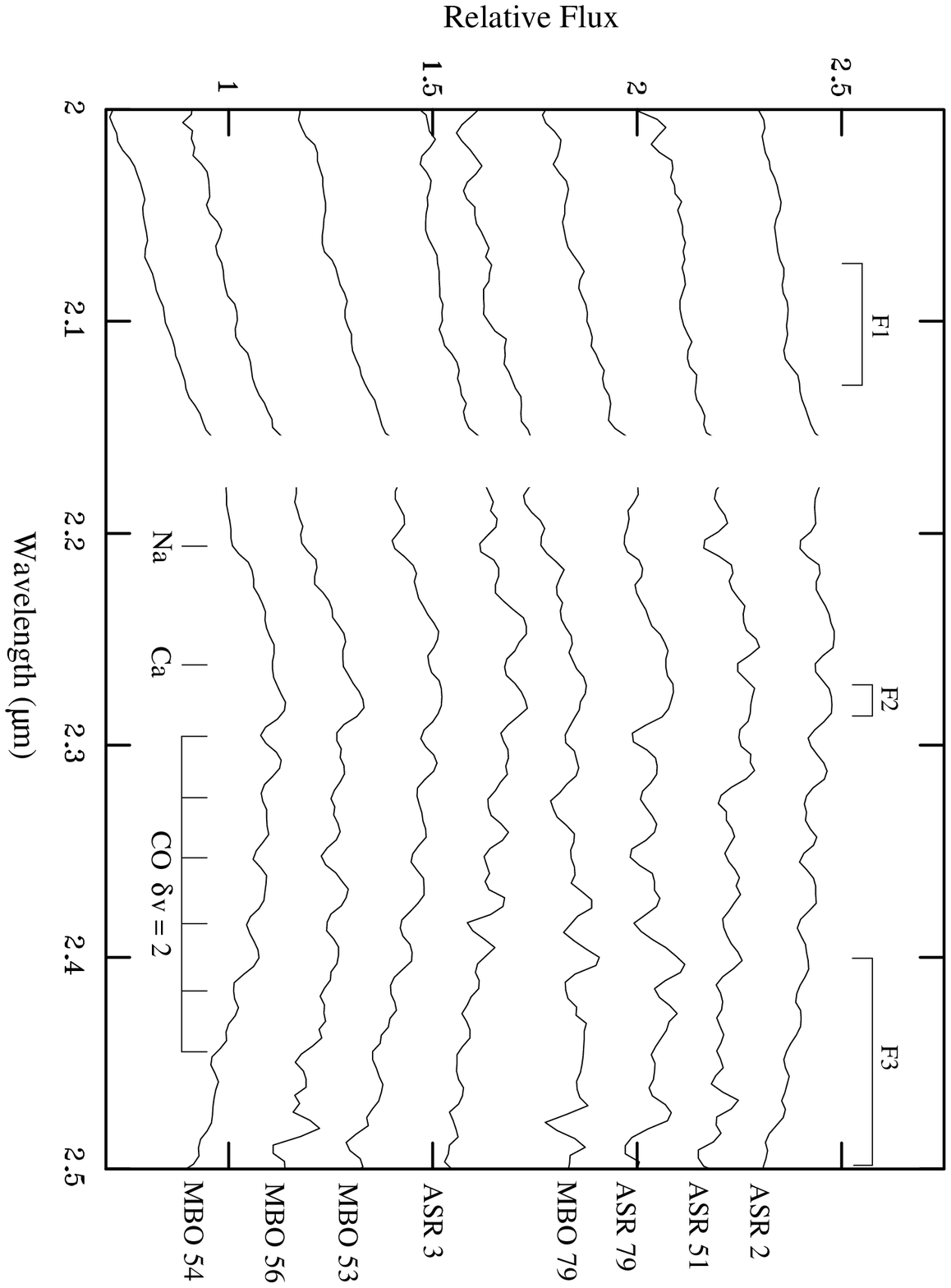}}
\newpage
\centerline{\includegraphics{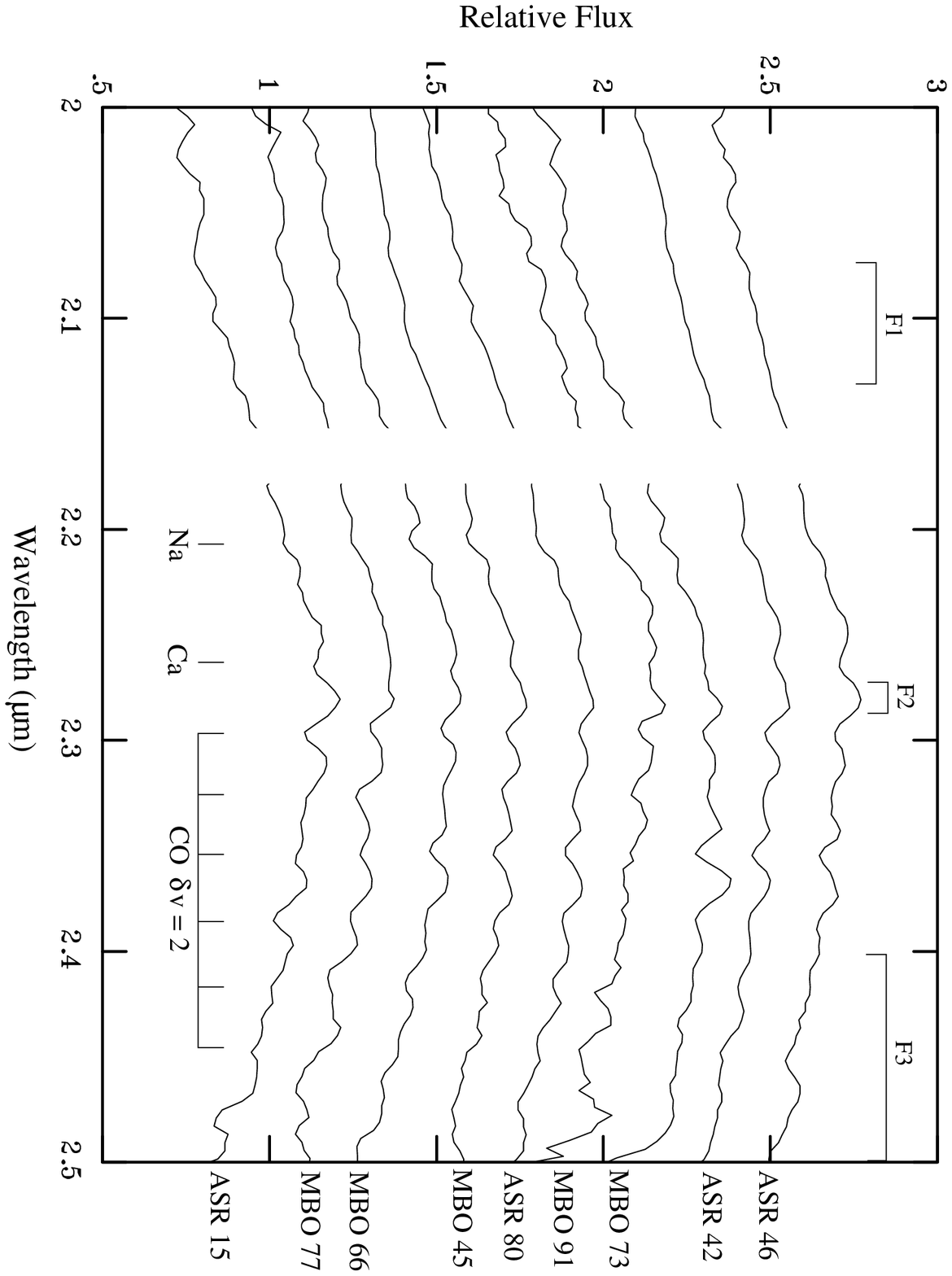}}
\newpage
\centerline{\includegraphics{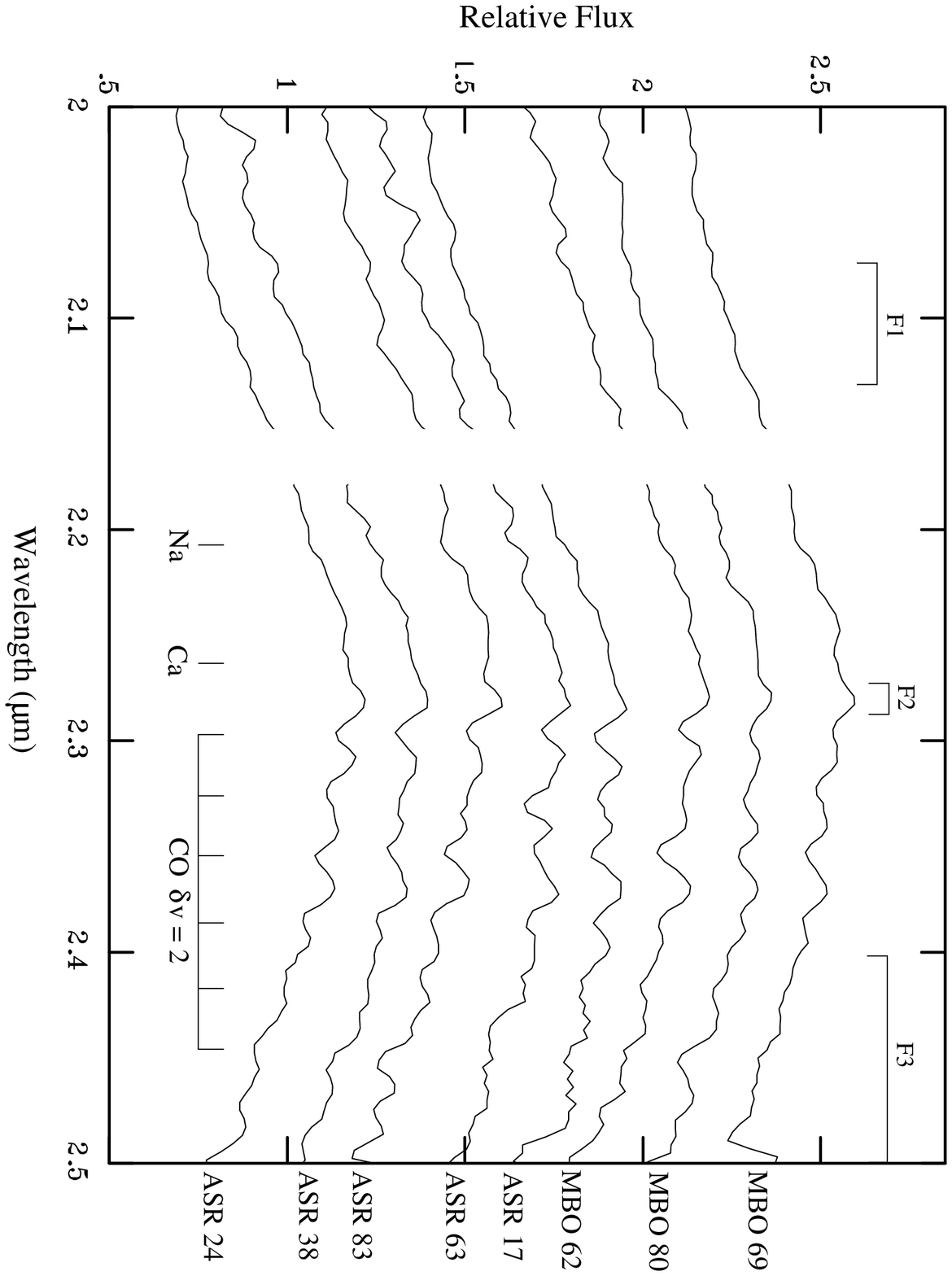}}
\newpage
\centerline{\includegraphics{Wilking.fig5.ps}}
\newpage
\centerline{\includegraphics{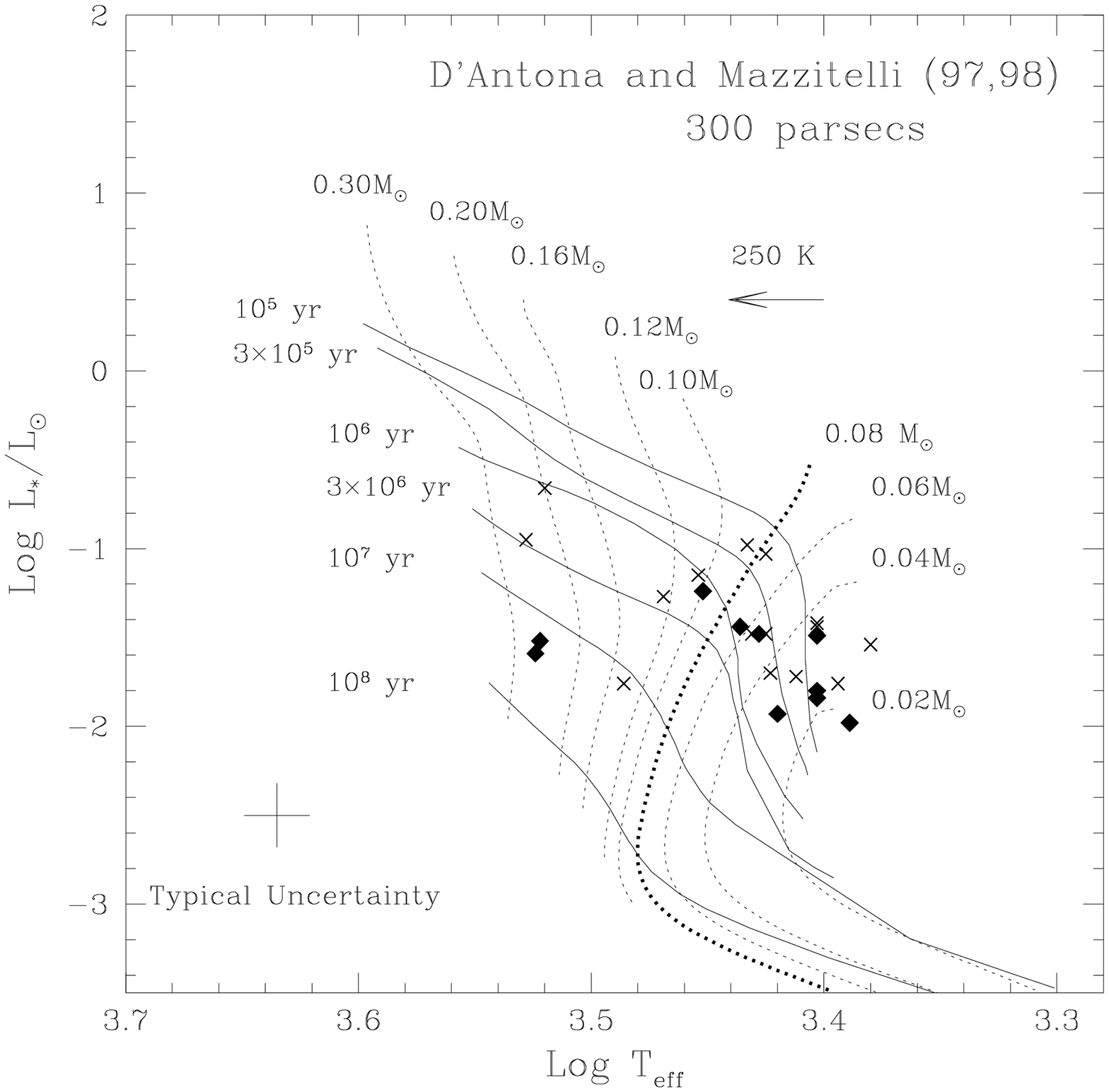}}
\newpage
\centerline{\includegraphics{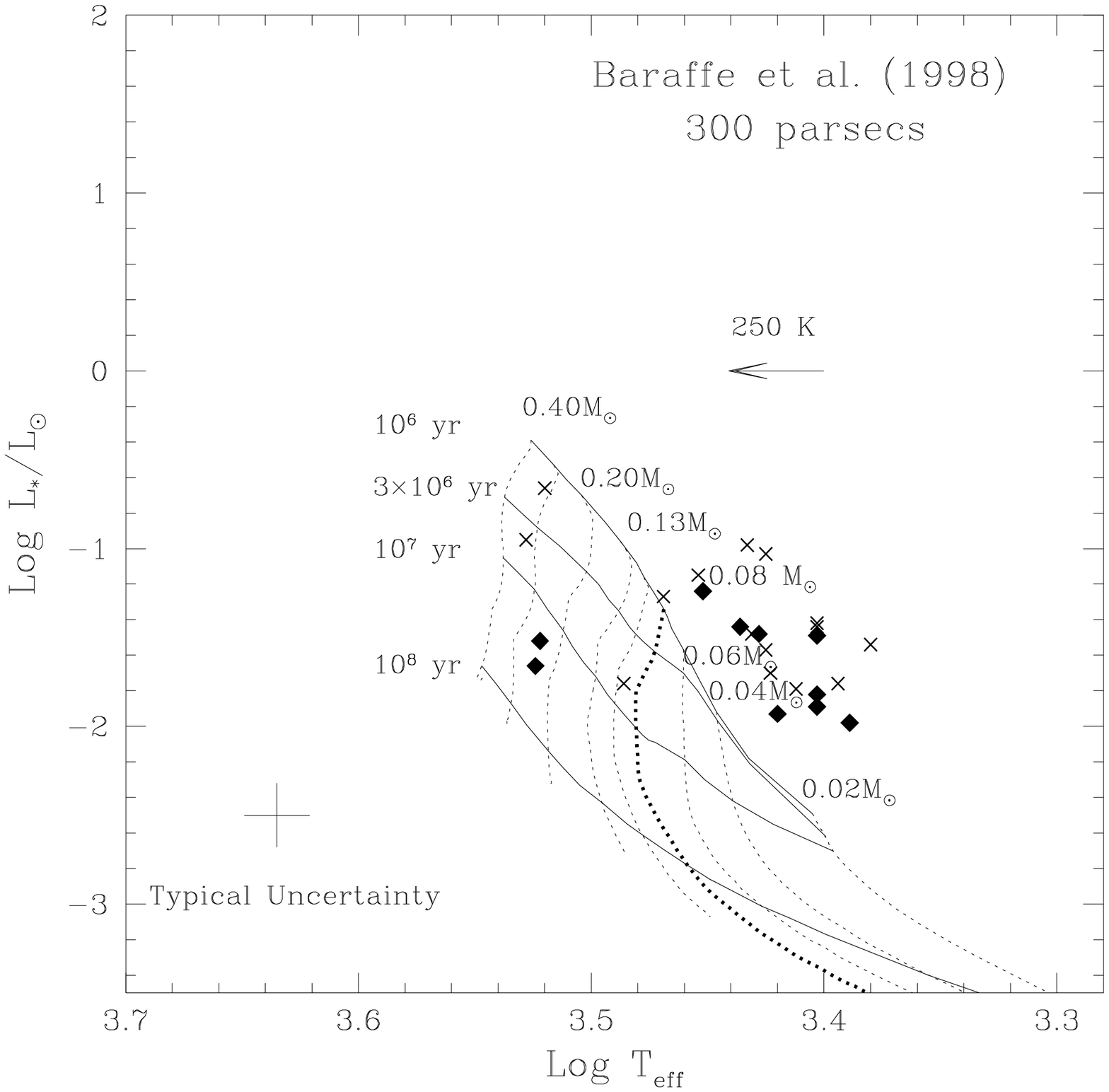}}
\newpage
\centerline{\includegraphics{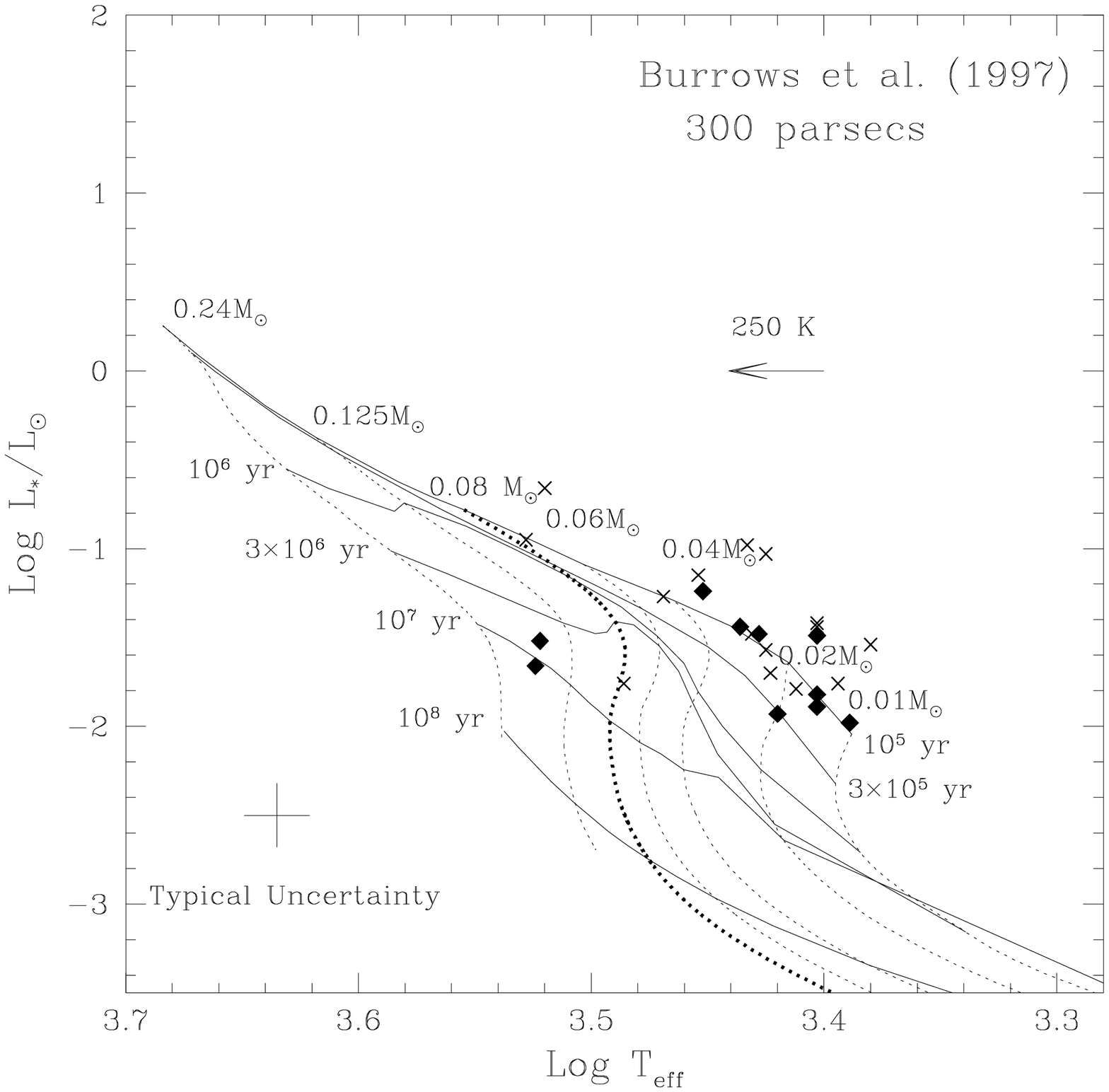}}
\newpage
\centerline{\includegraphics{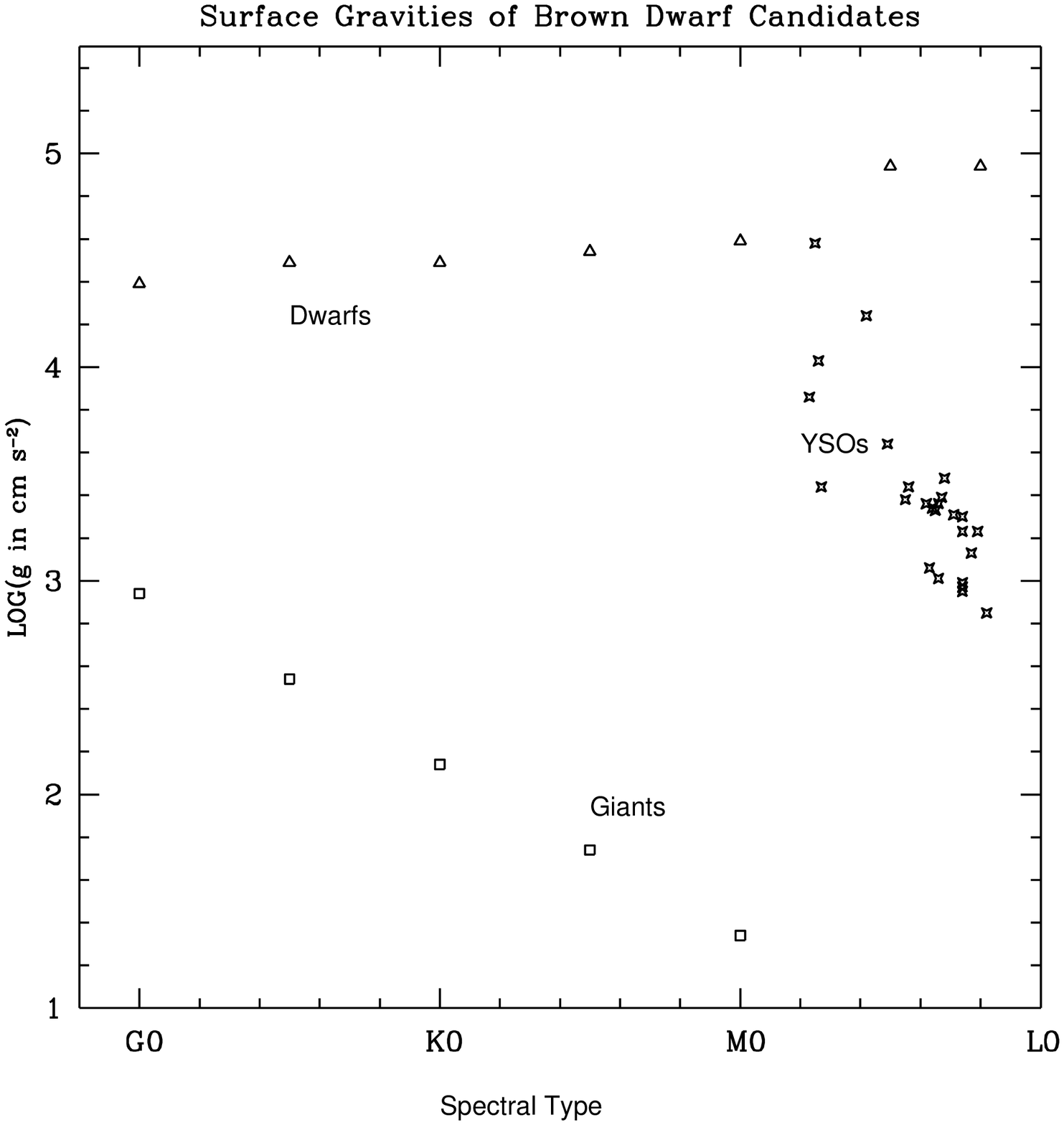}}
\newpage
\centerline{\includegraphics{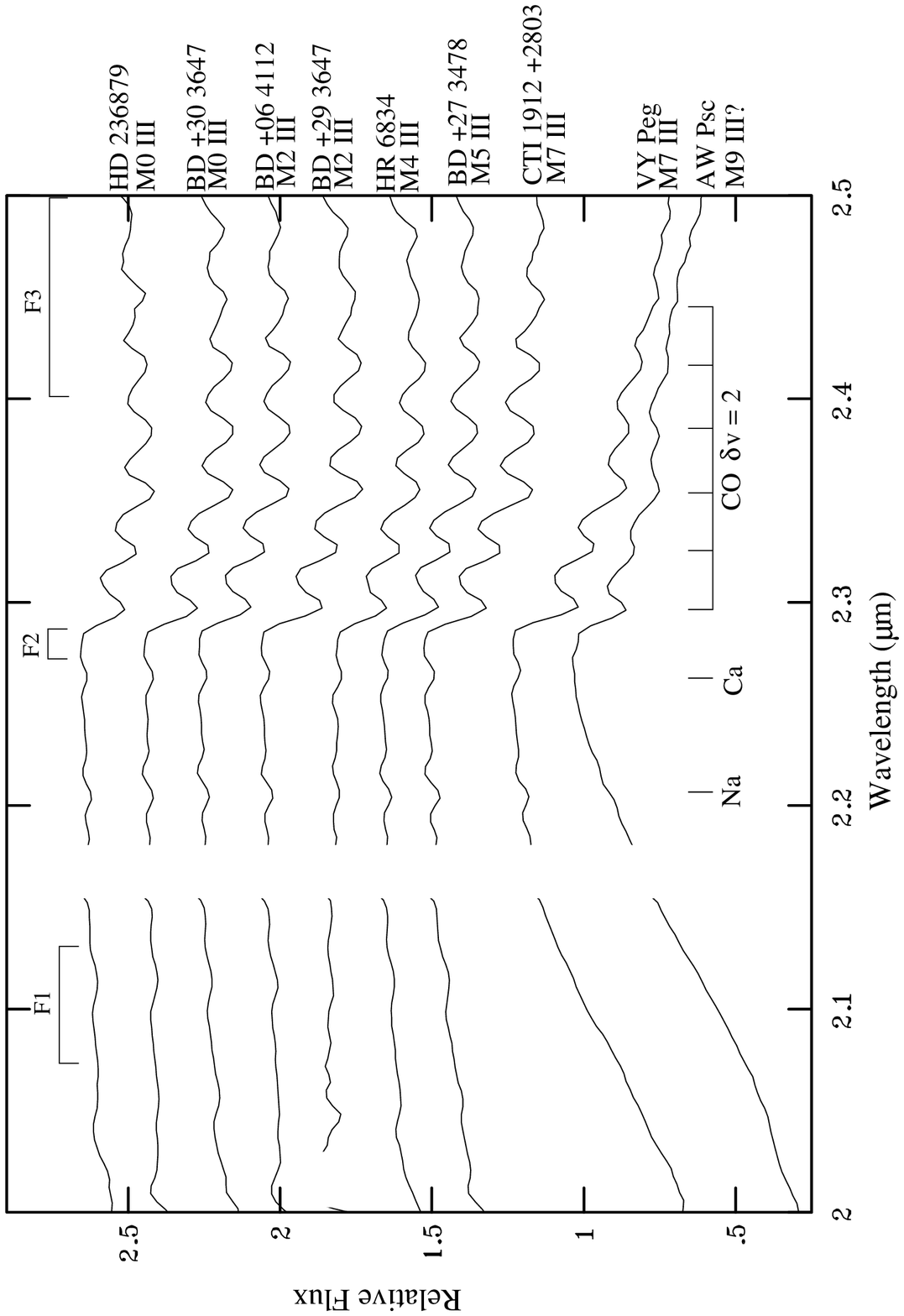}}

\end{document}